\documentclass[prd,showpacs,superscriptaddress,showkeys]{revtex4-2}
\usepackage{graphicx}
\usepackage{amsmath}
\usepackage{amssymb}
\usepackage{bm}
\usepackage{hyperref}
\usepackage{color}
\usepackage{booktabs}

\begin{document}

\title{Precise scaling relations for self-interacting bosonic dark matter stars}

\author{Zhaorui Zhang}
\affiliation{School of Space Science and Technology, Shandong University, Weihai 264209}
\author{Xudong Wang}
\email{xudongwang@mail.sdu.edu.cn}
\author{Bin Qi}
\email{bqi@sdu.edu.cn}
\affiliation{Shandong Provincial Key Laboratory of Nuclear Science, Nuclear 
Energy Technology and Comprehensive Utilization, Weihai Frontier Innovation 
Institute of Nuclear Technology, School of Nuclear Science, Energy and Power 
Engineering, Shandong University, Shandong 250061, China} 

\begin{abstract}
The structural properties of bosonic dark matter stars are systematically investigated, presenting precise scaling relations for the mass, radius, central density, and the properties of dark matter particles. The dark matter equation of state is derived from a complex scalar field theory with a quartic self-interaction potential $V(\phi) = \frac{\lambda}{4} |\phi|^4$, considering boson masses $m_{\phi}$ ranging from $10^{-9}$ to $10^{3}$ GeV and self-coupling constants $\lambda$ ranging from $0.01\pi$ to $100\pi$.
The scaling relation for the maximum mass of bosonic dark matter stars,  the corresponding critical radius and critical central density are obtained as
\[
M_{\text{max}} = 0.1 \frac{\sqrt{\lambda}}{m_\phi^2} M_\odot, \qquad R(M_{\text{max}}) = 0.9 \frac{\sqrt{\lambda}}{m_\phi^2} \ \text{km}, \qquad
\varepsilon_{\text{max}} = 2.1 \times 10^5 \frac{m_\phi^4}{\lambda} \ \mathrm{MeV/fm^3},
\]
where $m_\phi$ is in GeV, the relations for $R(M_{\text{max}})$ and $\varepsilon_{\text{max}}$  are  first put forward. The fitting relative error is less than $4\%$.
Based on these scaling relations, we further provide global analytical fits for the stable branch. The relationships between mass and central density as well as radius and central density can be described by a unified function of the form:
\[
\tilde{Y} = \frac{A}{\left[1 + \left(5\tilde{\varepsilon}\right)^h\right]^s},
\]
where for $Y=M$, $\tilde{M} \equiv M/M_{\text{max}}$, $A=1$, $h=-2$, $s=0.42$; for $Y=R$, $\tilde{R} \equiv R/R(M_{\text{max}})$, $A=1.634$, $h=1$, $s=0.28$; and $\tilde{\varepsilon} \equiv \varepsilon_0/\varepsilon_{\text{max}}$. The fitting relative error is less than $0.1\%$.
Furthermore, we find a simple quadratic polynomial mass-radius relation for bosonic dark matter stars.

The remarkable similarity of scaled stellar configurations across vastly different parameter ranges might imply an underlying mathematical structure. These analytical scaling relations establish direct connections between the macroscopic properties of dark matter stars and the microscopic properties of dark matter particles, enabling researchers to quickly obtain stellar configurations without repeated numerical calculations.
\end{abstract}

\maketitle

\section{Introduction}
\label{sec:intro}

Dark matter is one of the central issues in contemporary physics and astrophysics. A large body of astrophysical and cosmological observational evidence, ranging from galaxy rotation curves and cluster dynamics to anisotropies in the cosmic microwave background radiation, strongly indicates the existence of a component of matter that cannot be detected through standard electromagnetic radiation \cite{Bertone2005,Arcadi2018,Schumann2019,Clowe2006}. The latest observations from the Planck satellite precisely determine that dark matter accounts for approximately $26.4\%$ of the total energy density of the universe \cite{Planck2020}. However, the microscopic particle nature of dark matter remains an unsolved mystery. Although weakly interacting massive particles have long been considered the most attractive candidates, decades of direct and indirect detection experiments have yet to provide conclusive evidence for their existence \cite{Akerib2017,Aprile2018}. This situation has led the community to turn attention to other dark matter candidates, including axions, ultralight scalar fields, and primordial black holes \cite{Marsh2016,Hui2017,Green2021}.

Among the many alternative models, scalar field dark matter that can form macroscopic compact objects through gravitational binding—namely boson stars—has received widespread attention \cite{Kaup1968,Ruffini1969}. Wheeler first pointed out that a free complex scalar field without self-interaction can form gravitationally bound stable configurations \cite{Wheeler1955}. Subsequently, Kaup and Ruffini established the basic theoretical framework for boson stars \cite{Kaup1968,Ruffini1969}. For "mini-boson stars" composed of free complex scalar fields, the maximum mass is inversely proportional to the boson mass $m_\phi$, i.e., $M_{\text{max}} \sim 1/m_\phi$ \cite{Kaup1968}. Under typical particle mass scales, this mass is far smaller than the Chandrasekhar limit, the maximum mass achievable by fermionic stars supported by degeneracy pressure \cite{Chandrasekhar1931}, thus lacking astrophysical significance.
To overcome this limitation, self-interactions are introduced to generalize the form of the potential, providing additional pressure to resist gravitational collapse \cite{Schunck2003,Jetzer1992}. Colpi, Shapiro, and Wasserman systematically studied boson stars with a quartic self-interaction potential $V(\phi)=\lambda/4 |\phi|^4$ \cite{Colpi1986}. They found that repulsive self-interactions dominate the pressure in dense regions, leading to a scaling $M_{\text{max}} \sim \lambda^{1/2}/m_\phi^2$ for the maximum mass, which can reach or even exceed the mass scale of supermassive black holes when $\lambda$ is sufficiently large. They further pointed out that the maximum mass of such self-interacting boson stars can be compared to the Chandrasekhar mass of fermions with mass $m_f \sim \lambda^{-1/4}m_\phi$, and that self-interaction effects remain non-negligible even for $\lambda \ll 1$ \cite{Colpi1986}. In their classic review, Schunck and Mielke further noted that self-interactions not only increase the maximum mass but also endow boson stars with structured layered features—including an anisotropic scalar matter core, an exponentially decaying "halo," and a completely regular spacetime geometry—making them a potential new type of compact object. Giant boson stars can also contribute to galaxy rotation curves, providing a natural explanation for dark matter \cite{Schunck2003}. Liebling and Palenzuela systematically summarized various subtypes of boson stars and their dynamical properties, emphasizing that boson stars with repulsive self-interactions, due to their significantly enhanced dynamical stability, are widely used as dark matter sources and black hole mimickers, with rich phenomenological implications in high-energy astrophysics \cite{Liebling2012,Olivares2020,Vincent2021}. Recent studies have further confirmed that self-interactions can effectively suppress the dynamical instability of scalar boson stars \cite{Jaramillo2024}, while Bose-Einstein condensates with repulsive self-interactions can form stable solitonic structures on galactic scales, providing a microscopic foundation for dark matter models \cite{Chavanis2023}.

This paper focuses on the boson star model with a quartic self-interaction potential $V(\phi) = \frac{\lambda}{4} |\phi|^4$. Although the scaling relation $M_{\text{max}} \propto \sqrt{\lambda}/m_\phi^2$ for the maximum mass has been known since the work of Colpi et al. \cite{Colpi1986}, several important aspects remain unexplored. First, the numerical coefficients in these scaling relations have not been determined with high precision. Second, the scaling relation for the critical radius $R(M_{\text{max}})$ and its connection to the maximum mass have not been systematically quantified. Third, and most importantly, although previous work has pointed out the existence of scaling properties in boson stars \cite{Agnihotri2009}, high-precision analytical fitting formulas for the $M$--$\varepsilon_0$, $R$--$\varepsilon_0$, and $M$--$R$ relations over the entire stable branch have not yet appeared in the literature. The theoretical tool for studying boson star structure is the Tolman-Oppenheimer-Volkoff equation in the framework of general relativity \cite{Oppenheimer1939,Tooper1964}. However, numerically solving the TOV equations typically requires re-integration for each set of parameters $(m_\phi, \lambda)$, which hinders rapid theoretical predictions and astronomical data analysis. Therefore, finding universal analytical formulas describing the mass-radius-central density relations of boson stars has become an important task with significant practical value. Similar scaling relations have been explored in studies of neutron stars and quark stars \cite{Lattimer2001,Yagi2017}.

Through systematic scanning of a wide parameter space of $(m_\phi, \lambda)$, we obtain precise scaling relations for the maximum mass, critical radius, and critical central density of boson stars, and present for the first time unified analytical fits for the $M$--$\varepsilon_0$, $R$--$\varepsilon_0$, and $M$--$R$ relations over the entire stable branch. These analytical formulas allow researchers to quickly obtain stellar configurations for arbitrary parameter combinations without repeated numerical calculations, providing a convenient tool for subsequent astrophysical applications and observational constraints.

This paper is organized as follows. Section~\ref{sec:model} introduces the bosonic dark matter model and equation of state. Section~\ref{sec:eos} analyzes the behavior of the equation of state. Section~\ref{sec:bs} presents the numerical solutions of the TOV equations and the structure of boson stars. Section~\ref{sec:scaling} establishes the precise scaling relations at the extremal points. Section~\ref{sec:global_fit} presents global analytical fits for the $M$--$\varepsilon_0$, $R$--$\varepsilon_0$, and $M$--$R$ relations. Section~\ref{sec:sgra} applies these relations to constrain Sgr A*. Finally, Section~\ref{sec:SUMMARY} summarizes our findings and outlines future work.

\section{Model and Equation of State}
\label{sec:model}

\subsection{Bosonic Dark Matter Model}

We consider a bosonic dark matter model described by a complex scalar field $\phi$ with self-interaction. The Lagrangian density of the system is
\begin{equation}
\mathcal{L} = -\frac{1}{2}g^{\mu\nu}\phi^*_{;\mu}\phi_{;\nu} - \frac{1}{2}m_{\phi}^2|\phi|^2 - \frac{1}{4}\lambda|\phi|^4,
\label{eq:lagrangian}
\end{equation}
where $m_\phi$ is the boson mass and $\lambda$ is the dimensionless self-coupling constant. When $\lambda > 0$, the self-interaction is repulsive, favoring the formation of more massive compact objects \cite{Colpi1986}.

Under the mean-field approximation, the energy-momentum tensor of the system can be approximated as that of an ideal fluid, and anisotropies can be neglected \cite{Colpi1986}. Through mean-field treatment, the equation of state of the system can be obtained \cite{Chavanis2012}:
\begin{equation}
p = \frac{m_\phi^4c^5}{9\lambda\hbar^3}\left(\sqrt{1+\frac{3\lambda\hbar^3\varepsilon}{m_\phi^4c^5}}-1\right)^2.
\label{eq:eos_model}
\end{equation}

This equation of state exhibits two important asymptotic behaviors. In the low-energy density limit, $\frac{3\lambda\hbar^3\varepsilon}{m_\phi^4c^5} \ll 1$, Eq.~\eqref{eq:eos_model} can be approximated as
\begin{equation}
p \approx \frac{\lambda\hbar^3}{4m_\phi^4c^5}\varepsilon^2,
\label{eq:eos_low}
\end{equation}
exhibiting a soft quadratic behavior. In the high-energy density limit, $\frac{3\lambda\hbar^3\varepsilon}{m_\phi^4c^5} \gg 1$, Eq.~\eqref{eq:eos_model} approaches
\begin{equation}
p \approx \frac{\varepsilon}{3},
\label{eq:eos_high}
\end{equation}
the radiation-dominated linear equation of state.

\subsection{TOV Equation}

For static spherically symmetric compact objects, Einstein's field equations yield the Tolman-Oppenheimer-Volkoff equations describing hydrostatic equilibrium \cite{Tooper1964, Oppenheimer1939}:
\begin{align}
\frac{dp}{dr} &= -\frac{G(\varepsilon + p/c^2)(M + 4\pi r^3 p/c^2)}{r^2(1 - 2GM/c^2r)},
\label{eq:tov1} \\
\frac{dM}{dr} &= 4\pi r^2 \varepsilon/c^2,
\label{eq:tov2}
\end{align}
where $M(r)$ is the total mass enclosed within radius $r$, satisfying the boundary condition $M(0)=0$. The stellar surface is defined by the condition $p(R)=0$, and the total mass of the star is $M = M(R)$, with $R$ being the stellar radius.

We numerically solve Eqs.~\eqref{eq:tov1}-\eqref{eq:tov2} together with the equation of state \eqref{eq:eos_model} to compute the structure of boson stars for different parameters. To improve computational efficiency and cover a wide range of radial scales, we use a logarithmic grid from $r_{\text{min}} = 10^{-22}$ m to $r_{\text{max}} = 10^{22}$ m with $N = 20000$ grid points. The integration proceeds outward from the center and terminates when the pressure becomes negative (considered as reaching the stellar surface).

\section{Equation of State Behavior Analysis}
\label{sec:eos}

To comprehensively analyze the behavior of the equation of state \eqref{eq:eos_model}, we first study its variation with parameters. We select three typical self-coupling constants $\lambda = 0.01\pi, \pi, 100\pi$, covering the entire range from weak to strong coupling. For each $\lambda$, we calculate the pressure-energy density relation for different boson masses $m_\phi = 10^{-9}, 10^{-6}, 10^{-3}, 1, 10^{3}$ GeV.

\begin{figure}[htbp]
    \centering
    \includegraphics[width=\columnwidth]{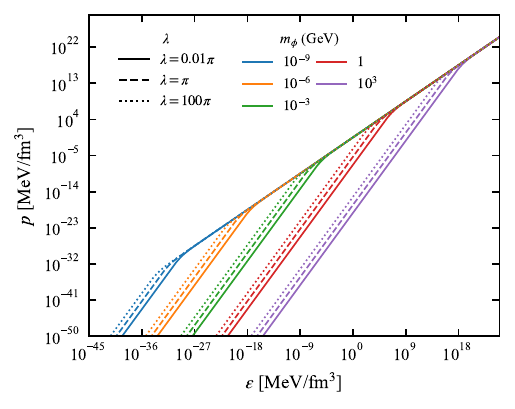}
    \caption{Equation of state for self-interacting bosonic dark matter with a quartic potential.
        The figure shows the pressure $p$ as a function of energy density $\varepsilon$ for different boson masses $m_\phi$ (colors) and self-coupling constants $\lambda$ (line styles).
        At low energy densities, the equation of state exhibits a soft quadratic behavior $p\propto\varepsilon^2$, while at high energy densities it asymptotes to the radiation-dominated limit $p=\varepsilon/3$.}
    \label{fig:eos}
\end{figure}

Figure~\ref{fig:eos} displays the equation of state, from which two important features can be observed:

First, for a fixed self-coupling constant $\lambda$, the equation of state becomes significantly softer as the boson mass $m_\phi$ increases. This means that at the same energy density, heavier bosons produce less pressure, making the resulting compact object more easily compressed. This behavior can be understood from the low-energy approximation in Eq.~\eqref{eq:eos_low}: the pressure is proportional to $m_\phi^{-4}$, so larger mass implies smaller pressure.

Second, for a fixed boson mass $m_\phi$, the equation of state in the low-energy density region becomes significantly stiffer as the self-coupling constant $\lambda$ increases, while in the high-energy density region, curves for different $\lambda$ gradually converge. This reflects the dual role of self-interaction: in the low-density region, repulsive self-interaction provides additional pressure support; in the high-density region, the radiation limit $p \approx \varepsilon/3$ dominates and is independent of $\lambda$.

\section{Structure of Boson Stars}
\label{sec:bs}

Having analyzed the equation of state behavior of self-interacting bosonic dark matter in the previous section, we now systematically study the structural properties of boson stars described by this equation of state by numerically solving the Tolman-Oppenheimer-Volkoff equations in general relativity. We focus on the $M$--$\varepsilon_0$ and $R$--$\varepsilon_0$ relations as well as the $M$--$R$ relation of boson stars.

Specifically, we choose boson masses $m_\phi = 10^{-9}, 10^{-6}, 10^{-3}, 1, 10^{3}$ GeV, covering the entire range from ultralight to heavy masses. The self-coupling constant $\lambda$ takes three typical values $0.01\pi$, $\pi$, and $100\pi$, corresponding to weak, moderate, and strong coupling regimes, respectively. For each set of parameters $(m_\phi, \lambda)$, we obtain a family of boson star solutions by varying the central energy density $\varepsilon_0$. For ease of expression, we refer to the radius $R(M_{\text{max}})$ and central energy density $\varepsilon_{\text{max}}$ corresponding to the maximum mass $M_{\text{max}}$ as the \textbf{critical radius} and \textbf{critical central density}, respectively.

\subsection{Boson Star Structure for Different Mass Parameters}

\begin{figure*}
    \centering
    \includegraphics[width=\textwidth]{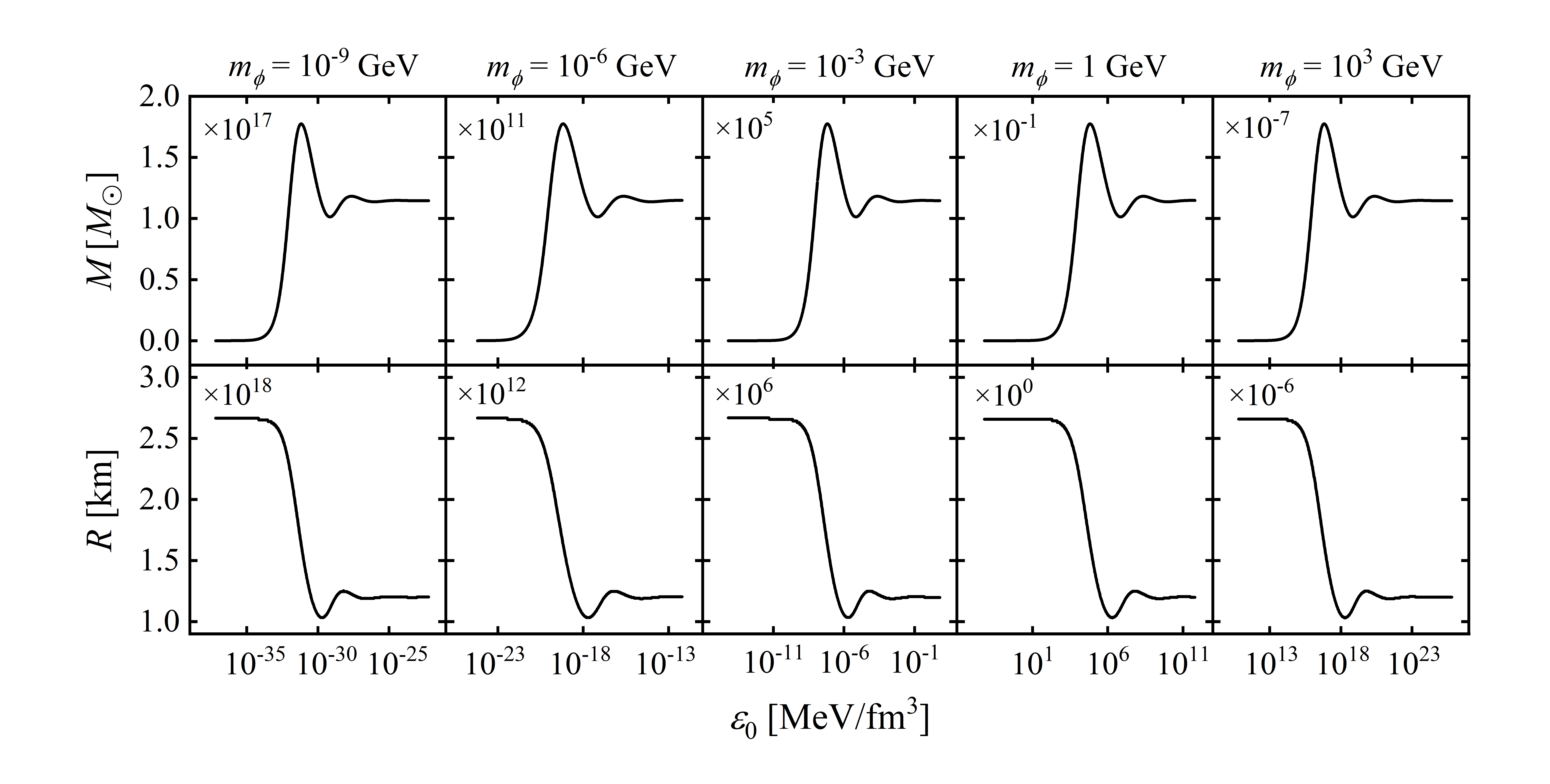}
    \caption{Fixed $\lambda=\pi$, variation of boson star mass and radius with central energy density for different $m_\phi$. From left to right: $m_\phi = 10^{-9},\ 10^{-6},\ 10^{-3},\ 1,\ 10^{3}$ GeV. Upper panels: $M$ (in $M_\odot$) as a function of $\varepsilon_0$; lower panels: $R$ (in km) as a function of $\varepsilon_0$, with $\varepsilon_0$ in units of MeV/fm$^3$.}
    \label{fig:combined_five}
\end{figure*}

First, we fix the self-coupling constant $\lambda=\pi$ and study the variation of boson star structure with the mass parameter $m_\phi$. Figure~\ref{fig:combined_five} shows the structural properties of boson stars for different mass parameters at $\lambda=\pi$, with each subfigure containing two panels displaying the $M$--$\varepsilon_0$ and $R$--$\varepsilon_0$ relations.

From the upper panels of Fig.~\ref{fig:combined_five} (the $M$--$\varepsilon_0$ relation), we see that each $M$--$\varepsilon_0$ curve has a maximum mass $M_{\text{max}}$. Before reaching the maximum, the mass increases monotonically with central density, corresponding to the stable equilibrium branch; after exceeding the maximum, the mass decreases with increasing central density and exhibits oscillations, corresponding to the unstable branch \cite{Harrison1965,Bardeen1966}. In this paper, the stable branch is defined as the region $\varepsilon_0 \le \varepsilon_{\text{max}}$, i.e., configurations where the central energy density does not exceed the critical central density $\varepsilon_{\text{max}}$. Configurations on the stable branch satisfy the standard radial perturbation hydrostatic stability criterion $\partial M/\partial \varepsilon_0 > 0$ \cite{Harrison1965}. This behavior is similar to that of ordinary neutron stars and is a universal feature of compact objects in general relativity, marking the limit of balance between gravity and repulsive forces (quantum pressure and self-interaction pressure).

The figure also shows that as the boson mass $m_\phi$ increases, the maximum mass decreases significantly and the corresponding critical central density shifts to higher values. For example, when $m_\phi = 10^{-9}$ GeV, the maximum mass can reach $M_{\text{max}} \sim 10^{17} M_\odot$; when $m_\phi = 10^{3}$ GeV, the maximum mass drops to $M_{\text{max}} \sim 10^{-7} M_\odot$, only of the order of Earth's mass. This behavior can be understood from the low-energy approximation of the equation of state: in the low-density region, $p \propto \lambda \varepsilon^2/m_\phi^4$, the pressure is proportional to $m_\phi^{-4}$, so heavier particles produce less pressure and consequently support smaller stellar masses.

Next, examine the lower panels of Fig.~\ref{fig:combined_five} (the $R$--$\varepsilon_0$ relation). The radius at the maximum mass on each $R$--$\varepsilon_0$ curve is denoted $R(M_{\text{max}})$, i.e., the critical radius. In the low-density region, the radius decreases with increasing central density; in the high-density region, the radius remains nearly constant after small oscillations.

The figure also shows that as $m_\phi$ increases, the critical radius $R(M_{\text{max}})$ decreases significantly. For $m_\phi = 10^{-9}$ GeV, the maximum radius can reach $R(M_{\text{max}}) \sim 10^{18}$ km, while for $m_\phi = 10^{3}$ GeV, the critical radius is only $R(M_{\text{max}}) \sim 10^{-6}$ km.

\begin{figure*}
    \centering
    \includegraphics[width=\textwidth]{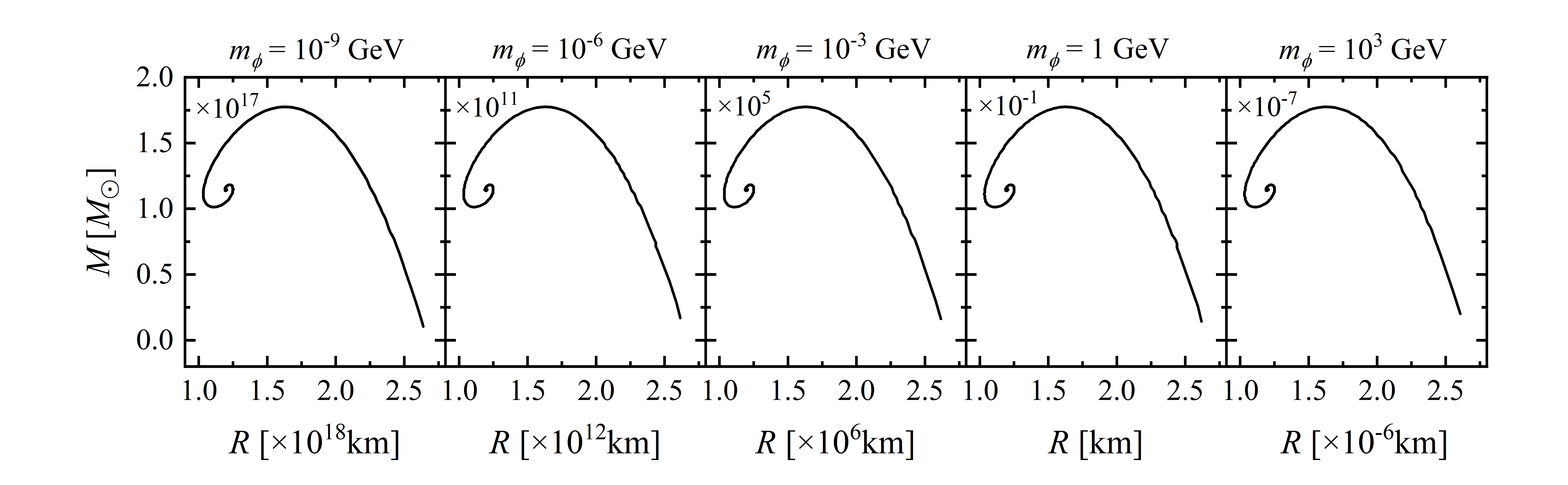}
    \caption{Fixed self-coupling constant $\lambda=\pi$, mass-radius relation of boson stars for different boson masses $m_\phi$. From left to right: $m_\phi = 10^{-9},\ 10^{-6},\ 10^{-3},\ 1,\ 10^{3}$ GeV. Horizontal axis: radius $R$ in km; vertical axis: mass $M$ in solar masses $M_\odot$. Each curve is obtained by varying the central energy density $\varepsilon_0$.}
    \label{fig:mr_combined}
\end{figure*}

Figure~\ref{fig:mr_combined} shows the mass-radius relation of boson stars for the same parameters, which is one of the most direct observables of compact objects and a key bridge connecting theoretical models with astronomical observations \cite{Lattimer2001}.

An important characteristic quantity can be extracted from Fig.~\ref{fig:mr_combined}—the critical radius $R(M_{\text{max}})$. For $m_\phi = 10^{-9}$ GeV, the maximum mass can reach $M_{\text{max}} \sim 10^{17} M_\odot$, with a corresponding critical radius of approximately $R(M_{\text{max}}) \sim 10^{18}$ km. This indicates that ultralight bosons can form supermassive compact objects, providing theoretical support for the dark matter origin hypothesis of supermassive compact objects such as Sgr A* at the Galactic center.

\subsection{Boson Star Structure for Different Coupling Constants}

\begin{figure*}
    \centering
    \includegraphics[width=\textwidth]{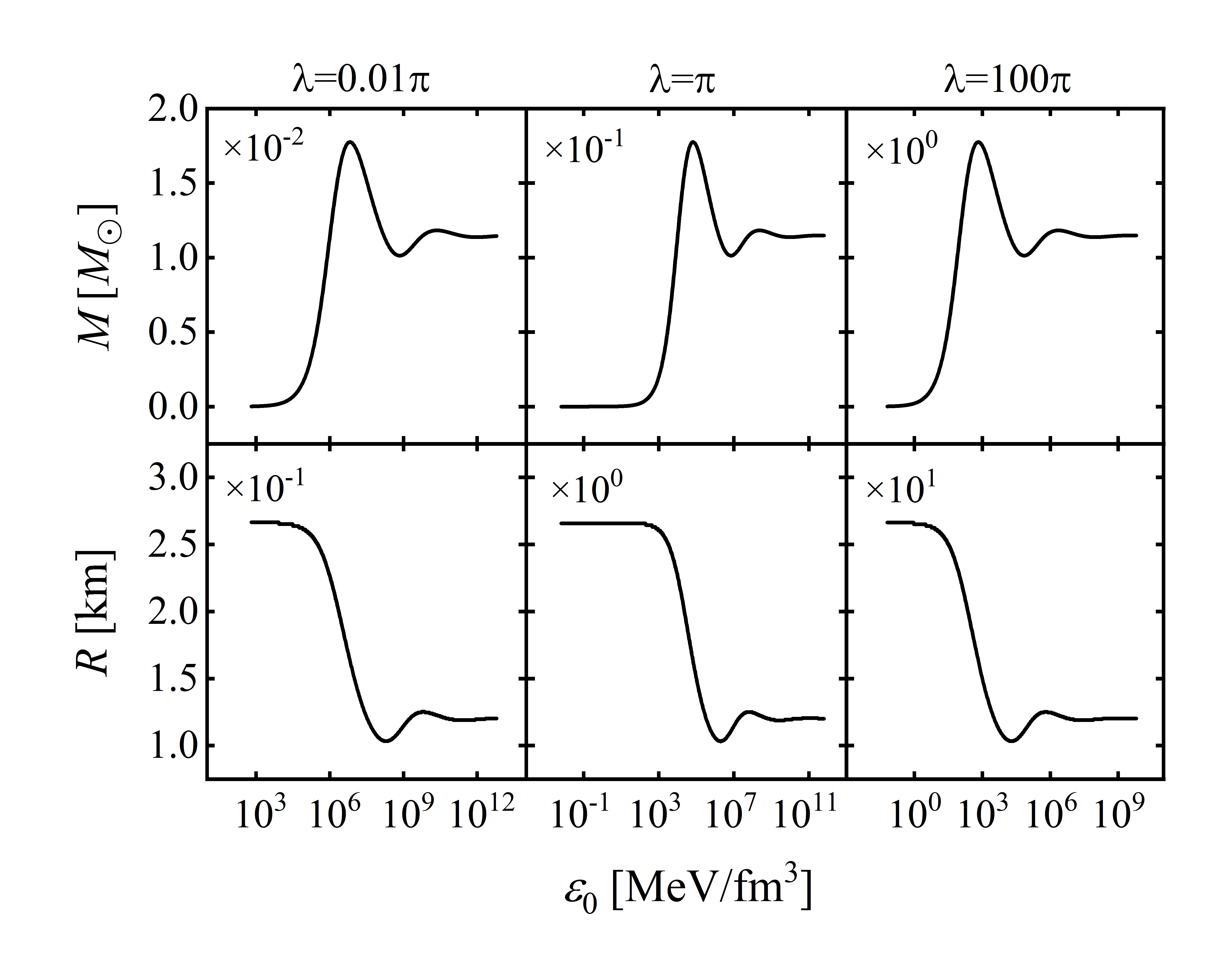}
    \caption{Fixed mass $m_\phi = 1$ GeV, variation of boson star mass and radius with central energy density for different self-coupling constants $\lambda$. From left to right: $\lambda = 0.01\pi,\ \pi,\ 100\pi$. Upper panels: $M$ (in $M_\odot$) as a function of $\varepsilon_0$; lower panels: $R$ (in km) as a function of $\varepsilon_0$, with $\varepsilon_0$ in units of MeV/fm$^3$.}
    \label{fig:combined_five_lambda}
\end{figure*}

\begin{figure*}
    \centering
    \includegraphics[width=\textwidth]{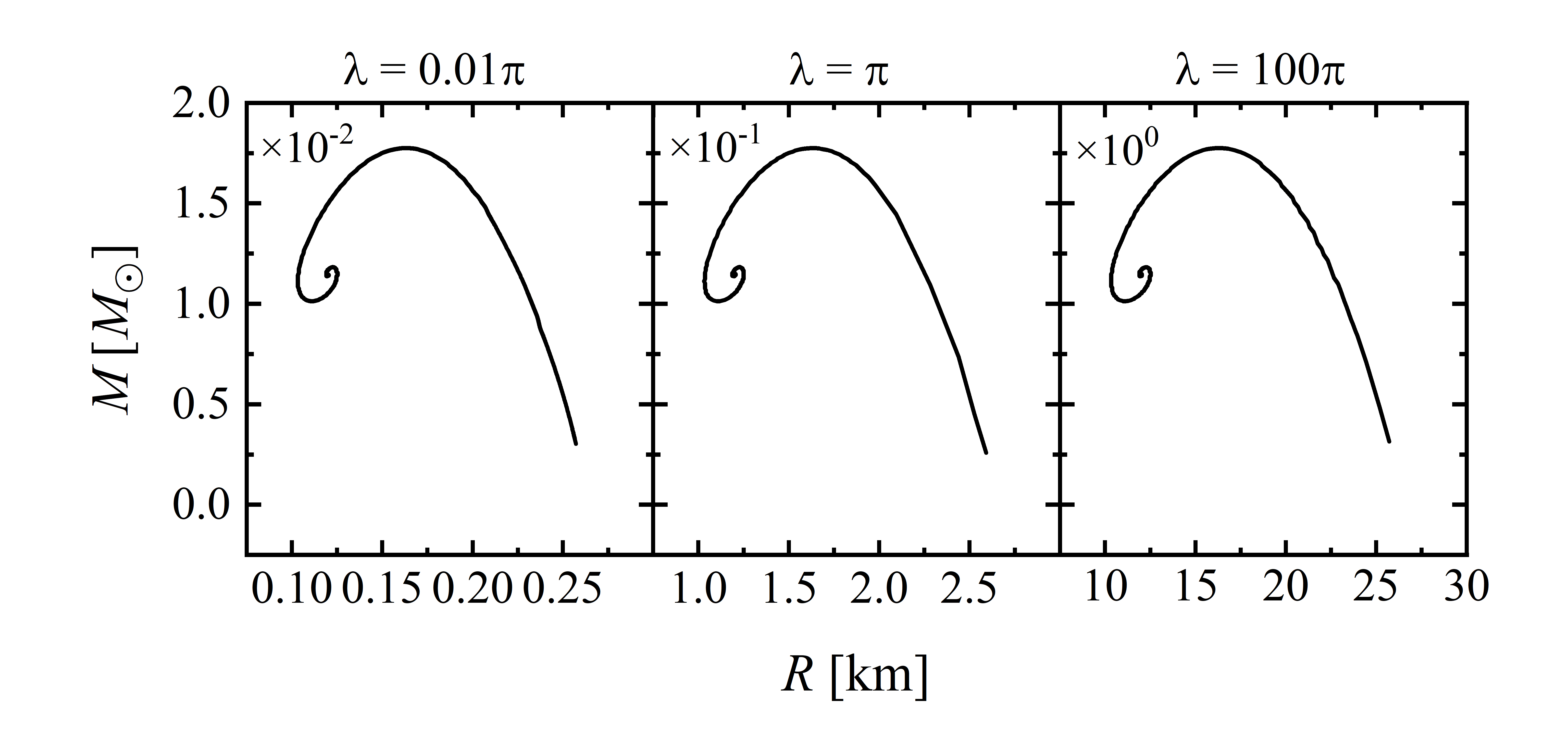}
    \caption{Fixed mass $m_\phi = 1$ GeV, mass-radius relation of boson stars for different self-coupling constants $\lambda$. From left to right: $\lambda = 0.01\pi,\ \pi,\ 100\pi$. Horizontal axis: radius $R$ in km; vertical axis: mass $M$ in solar masses $M_\odot$. Each curve is obtained by varying the central energy density $\varepsilon_0$.}
    \label{fig:mr_combined_lambda}
\end{figure*}

Next, we fix the mass parameter $m_\phi = 1$ GeV and study the variation of boson star structure with the self-coupling constant $\lambda$. Figure~\ref{fig:combined_five_lambda} shows the $M$--$\varepsilon_0$ and $R$--$\varepsilon_0$ relations for different self-coupling constants, and Fig.~\ref{fig:mr_combined_lambda} shows the corresponding mass-radius relations.

From the upper panels of Fig.~\ref{fig:combined_five_lambda} (the $M$--$\varepsilon_0$ relation), we see that as $\lambda$ increases, the maximum mass increases significantly and the corresponding critical central density shifts to lower values. This indicates that stronger self-interaction provides greater repulsive pressure, thus supporting larger stellar masses.

From the lower panels of Fig.~\ref{fig:combined_five_lambda} (the $R$--$\varepsilon_0$ relation), we observe that as $\lambda$ increases, the critical radius $R(M_{\text{max}})$ also increases accordingly. At $\lambda = 0.01\pi$, the critical radius is approximately $0.17$ km, while at $\lambda = 100\pi$, it reaches about $17$ km.

From the mass-radius relations in Fig.~\ref{fig:mr_combined_lambda}, we find that as $\lambda$ increases, the $M$--$R$ curves shift upward and rightward, indicating that strongly coupled boson stars have larger radii for the same mass. In the weak coupling regime, the maximum mass of boson stars is relatively small, $M_{\text{max}} \approx 1.77\times10^{-2} M_\odot$; under moderate coupling, the maximum mass increases to $M_{\text{max}} \approx 0.17 M_\odot$; in the strong coupling regime, the maximum mass further increases to $M_{\text{max}} \approx 1.7 M_\odot$.

\section{Precise Scaling Relations at Extremal Points}
\label{sec:scaling}

\subsection{Scaling Relation for the Maximum Mass}

\begin{figure}[htbp]
    \centering
    \includegraphics[width=0.6\textwidth]{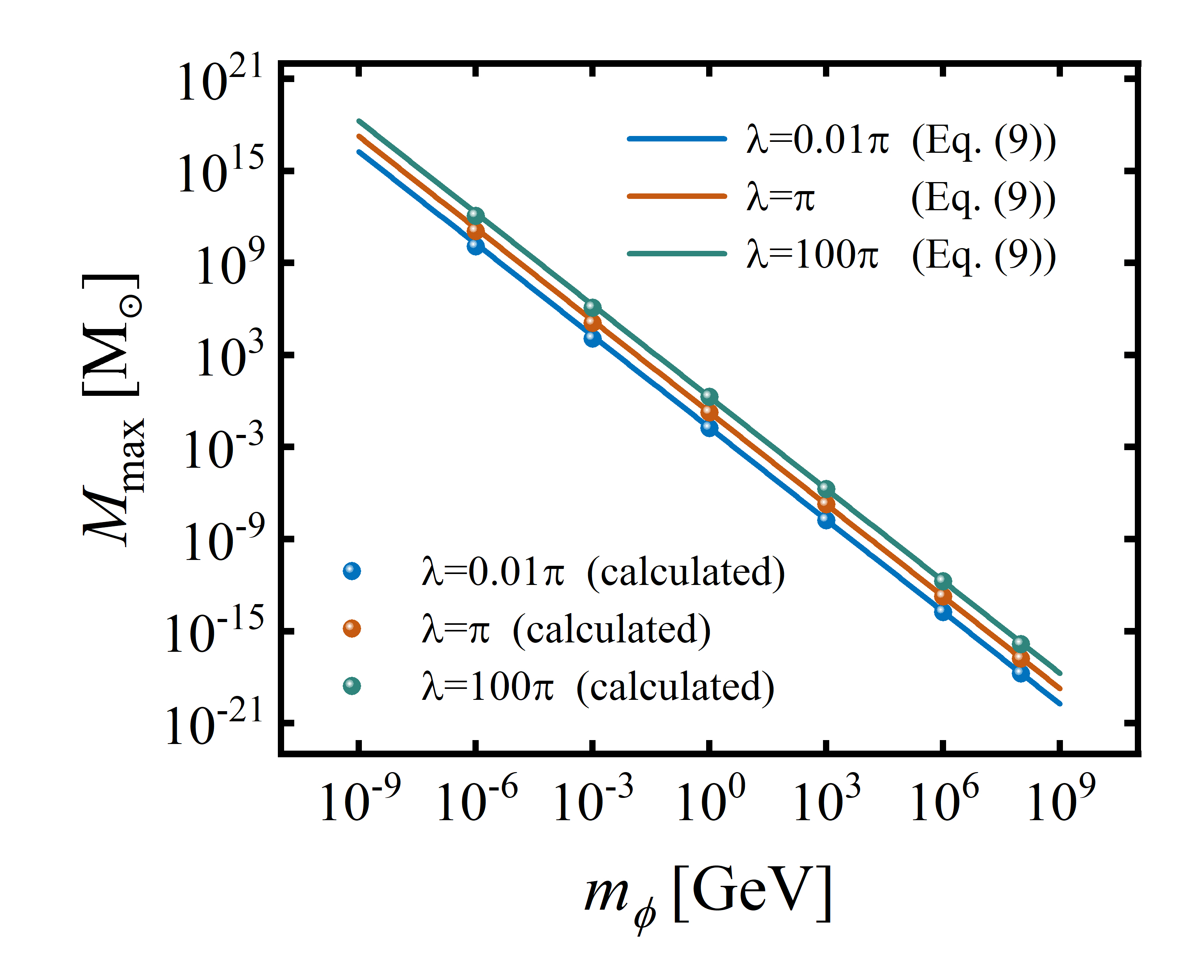}
    \caption{Fixed self-coupling constant $\lambda$, variation of maximum mass $M_{\text{max}}$ with boson mass $m_\phi$. Solid lines represent the theoretical scaling relation Eq.~\eqref{eq:Mmax_scaling}, where blue, red, and green correspond to $\lambda = 0.01\pi$, $\pi$, and $100\pi$, respectively. Discrete points are numerical results from solving the TOV equations.}
    \label{fig:Mmax_m}
\end{figure}

Through systematic analysis of the maximum mass for different parameters, we discover precise scaling relations. Figure~\ref{fig:Mmax_m} shows the variation of the maximum mass with boson mass for fixed self-coupling constant $\lambda$, where the solid line represents the theoretical scaling relation $M_{\text{max}} = 0.1\sqrt{\lambda}/m_\phi^2$, and the discrete points are numerical results from solving the TOV equations.

Figure~\ref{fig:Mmax_m} clearly shows perfect agreement between numerical points and theoretical curves. On a log-log scale, $M_{\text{max}}$ exhibits a strict power-law relation with $m_\phi$, with a fitted slope of approximately $-2.00 \pm 0.01$, i.e.,
\begin{equation}
M_{\text{max}} \propto m_\phi^{-2}.
\label{eq:scaling_m}
\end{equation}
This scaling relation can be understood from the low-energy asymptotic form of the equation of state: when $\varepsilon$ is small, $p \propto \lambda \varepsilon^2/m_\phi^4$, and dimensional analysis yields $M \propto m_\phi^{-2}$ \cite{Colpi1986}.

Similarly, fixing the boson mass $m_\phi$ and varying the self-coupling constant $\lambda$ in numerical calculations, we find that the maximum mass $M_{\text{max}}$ is proportional to $\sqrt{\lambda}$, i.e.,
\begin{equation}
M_{\text{max}} \propto \lambda^{1/2}.
\label{eq:scaling_lambda}
\end{equation}
This behavior reflects the supporting role of repulsive self-interaction on stellar structure: the larger $\lambda$ is, the greater the additional pressure provided, and the larger the maximum mass that can be supported.

Combining these two scaling relations, the maximum mass can be uniformly expressed as
\begin{equation}
M_{\text{max}} = 0.1 \frac{\sqrt{\lambda}}{m_\phi^2},
\label{eq:Mmax_scaling}
\end{equation}
with $m_\phi$ in units of GeV and $M_{\text{max}}$ in solar masses $M_\odot$.

\subsection{Scaling Relation for the Critical Radius}

\begin{figure}[htbp]
    \centering
    \includegraphics[width=0.6\textwidth]{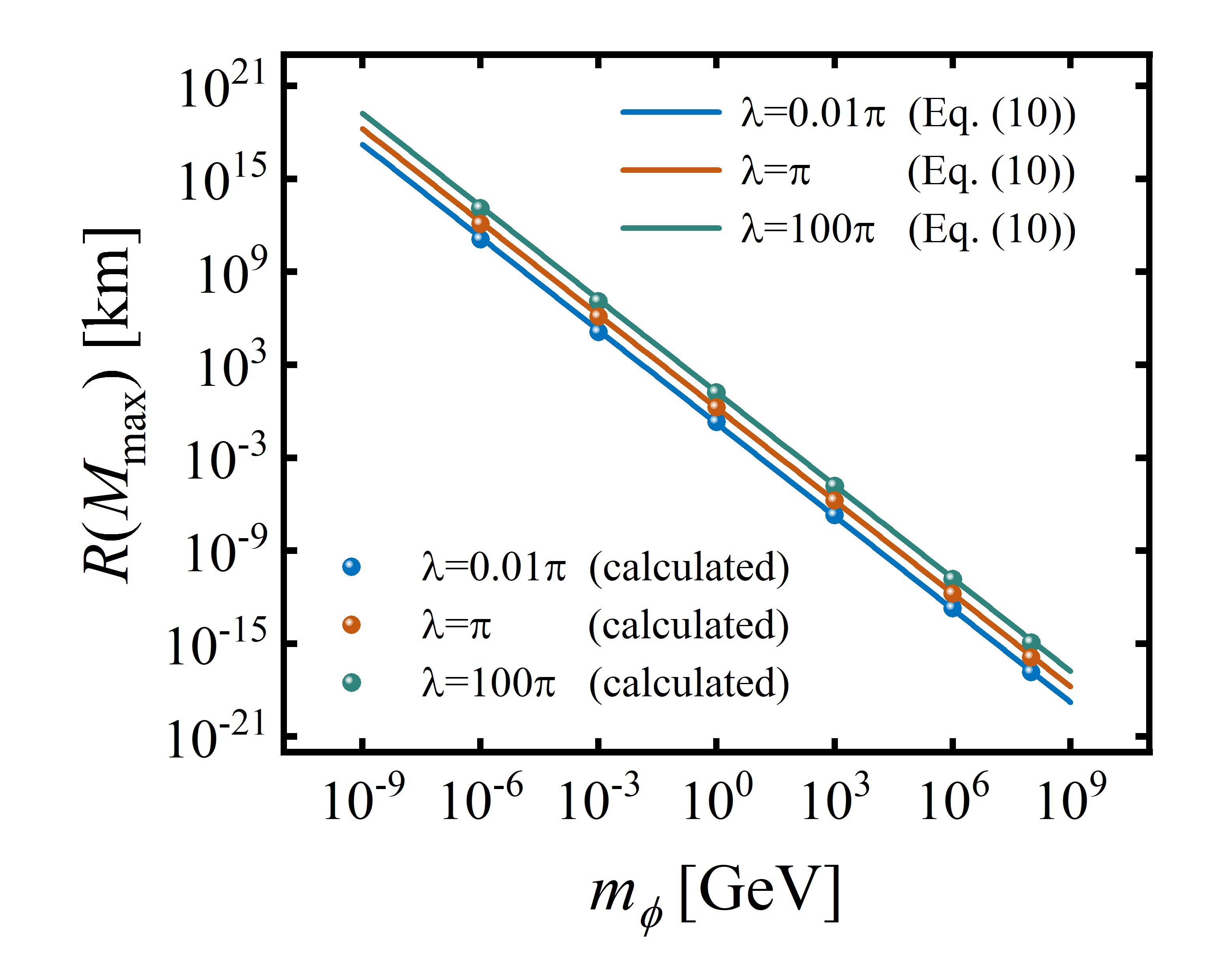}
    \caption{Fixed self-coupling constant $\lambda$, variation of critical radius $R(M_{\text{max}})$ with boson mass $m_\phi$. Solid lines represent the theoretical scaling relation Eq.~\eqref{eq:Rmax_scaling}, with different colors corresponding to different $\lambda$ values. Discrete points are numerical results from solving the TOV equations.}
    \label{fig:Rmax_m}
\end{figure}

For compact object research, the critical radius $R(M_{\text{max}})$ is another important physical quantity. We systematically analyze this quantity as well. Figure~\ref{fig:Rmax_m} shows the variation of $R(M_{\text{max}})$ with $m_\phi$ for fixed $\lambda$, with solid lines representing the theoretical scaling relation and discrete points being TOV numerical solutions.

The analysis results show that the critical radius $R(M_{\text{max}})$ exhibits exactly the same scaling behavior with $m_\phi$ and $\lambda$ as the maximum mass $M_{\text{max}}$:
\begin{equation}
R(M_{\text{max}}) = 0.9 \frac{\sqrt{\lambda}}{m_\phi^2},
\label{eq:Rmax_scaling}
\end{equation}
with $R(M_{\text{max}})$ in km. It is noteworthy that $M_{\text{max}}$ and $R(M_{\text{max}})$ have identical dependencies on $m_\phi$ and $\lambda$, reflecting the self-similarity of compact stellar structures.

From Eqs.~\eqref{eq:Mmax_scaling} and \eqref{eq:Rmax_scaling}, the ratio of the maximum mass to the critical radius for the same $(m_\phi, \lambda)$ combination is a universal constant:
\begin{equation}
\frac{M_{\text{max}}}{R(M_{\text{max}})} = \frac{1}{9}  \frac{\ M_\odot}{\text{km}}.
\label{eq:mass_radius_ratio}
\end{equation}
That is, $M_{\text{max}} = \frac{1}{9} R(M_{\text{max}})$. Linear fitting of the data yields a slope of $1.00006 \pm 0.00004$ ($R^2 > 0.9999$, where $R^2$ denotes the coefficient of
determination), confirming $M_{\text{max}} \propto R(M_{\text{max}})$ and the ratio $M_{\text{max}}/R(M_{\text{max}}) \approx \frac{1}{9}$, with error less than $2\%$.

From the scaling relations, we can further obtain the compactness of the boson star at the critical point:
\begin{equation}
C \equiv \frac{GM_{\text{max}}}{R(M_{\text{max}})c^2} = \frac{G}{c^2}\cdot\frac{1}{9}\,\frac{M_\odot}{\text{km}} = 0.164.
\end{equation}
This value is comparable in magnitude to the compactness of neutron stars, indicating that boson stars with quartic self-interaction are also highly compact at the maximum mass.

\subsection{Scaling Relation for the Critical Central Density}

The critical central density $\varepsilon_{\text{max}}$ also exhibits universal scaling behavior. Starting from the low-energy approximation of the equation of state $p \propto \lambda\varepsilon^2/m_\phi^4$, dimensional analysis gives
\begin{equation}
\varepsilon_{\text{max}} = 2.1 \times 10^5 \frac{m_\phi^4}{\lambda},
\label{eq:epsm_scaling}
\end{equation}
where $\varepsilon_{\text{max}}$ is in MeV/fm$^3$ and $m_\phi$ is in GeV, $R^2 > 0.9999$. This scaling relation is consistent with those for $M_{\text{max}}$ and $R(M_{\text{max}})$: from Eqs.~\eqref{eq:Mmax_scaling}, \eqref{eq:Rmax_scaling}, and \eqref{eq:epsm_scaling}, we can verify $M_{\text{max}} \propto \sqrt{\lambda}/m_\phi^2 \propto 1/\sqrt{\varepsilon_{\text{max}}}$, consistent with the universal scaling law for compact stars.

These three scaling relations form the theoretical foundation for subsequent studies of the macroscopic properties of boson stars and their comparison with astronomical observations.

\section{Global Analytical Fits}
\label{sec:global_fit}

\subsection{\texorpdfstring{$M$--$\varepsilon_0$}{M-epsilon0} and \texorpdfstring{$R$--$\varepsilon_0$}{R-epsilon0} Relations}

In Section~\ref{sec:scaling}, we established precise scaling relations for the maximum mass, critical radius, and critical central density. Based on these, we now present global analytical fits for the $M$--$\varepsilon_0$ and $R$--$\varepsilon_0$ relations over the entire stable branch. The stable branch is defined as configurations with central energy density not exceeding the critical central density $\varepsilon_{\text{max}}$, satisfying the standard stability criterion $\partial M/\partial \varepsilon_0 > 0$ \cite{Harrison1965}. We find that these two relations can be described by a unified function of the form:
\begin{equation}
\tilde{Y} = \frac{A}{\left[1 + \left({5\tilde{\varepsilon}}\right)^h\right]^s},
\label{eq:unified_fit}
\end{equation}
where for $Y=M$, $\tilde{M} \equiv M/M_{\text{max}}$, $A=1$, $h=-2$, $s=0.42$; for $Y=R$, $\tilde{R} \equiv R/R(M_{\text{max}})$, $A=1.634$, $h=1$, $s=0.28$; and $\tilde{\varepsilon} \equiv \varepsilon_0/\varepsilon_{\text{max}}$. These parameters remain constant over the entire parameter range $10^{-9} \le m_\phi/\text{GeV} \le 10^{3}$ and $0.01\pi \le \lambda \le 100\pi$. The fitting error of Eq.~\eqref{eq:unified_fit} over the entire stable branch is less than $0.1\%$ with $R^2 > 0.9995$.

It should be noted that the parameters $A$, $h$, and $s$ in the fitting formula are shape parameters whose numerical values do not carry independent physical significance. The precise maximum mass, critical radius, and critical central density should be calculated using Eqs.~\eqref{eq:Mmax_scaling}, \eqref{eq:Rmax_scaling}, and \eqref{eq:epsm_scaling}, while Eq.~\eqref{eq:unified_fit} has the advantage of describing the entire stable branch with a unified dimensionless functional form. Based on the power-law form of the scaling relations and the self-similarity of the TOV equations, these fitting formulas can be reasonably extrapolated to other $(m_\phi, \lambda)$ combinations beyond the parameter range considered in this paper.

Several important physical features can be read from the fitting results:

\begin{enumerate}
    \item Scaling relation for characteristic density: $\varepsilon_{\text{max}} \propto m_\phi^4/\lambda$, together with $Y_{\text{max}} \propto \sqrt{\lambda}/m_\phi^2$, reflects the self-similarity of the system.
    \item Difference in shape parameters: for the mass fit $h=-2<0$ while for the radius fit $h=1>0$, indicating that the contraction behavior of radius with central density and the growth behavior of mass have different functional characteristics, consistent with the trends observed in Figs.~\ref{fig:combined_five} and \ref{fig:combined_five_lambda}.
    \item Variation of radius within the stable branch: for the radius fit, within the stable branch ($\varepsilon_0 \le \varepsilon_{\text{max}}$), when $\varepsilon_0 \ll \varepsilon_{\text{max}}$, $\tilde{R} \approx A = 1.634$, i.e., $R \approx 1.634 R(M_{\text{max}})$; at the critical point, $\tilde{R} = 1$, i.e., $R = R(M_{\text{max}})$. The radius varies by a factor of about $1.6$, quantitatively characterizing the evolution of the boson star radius with increasing central density.
\end{enumerate}

\begin{figure}[htbp]
    \centering
    \includegraphics[width=0.8\columnwidth]{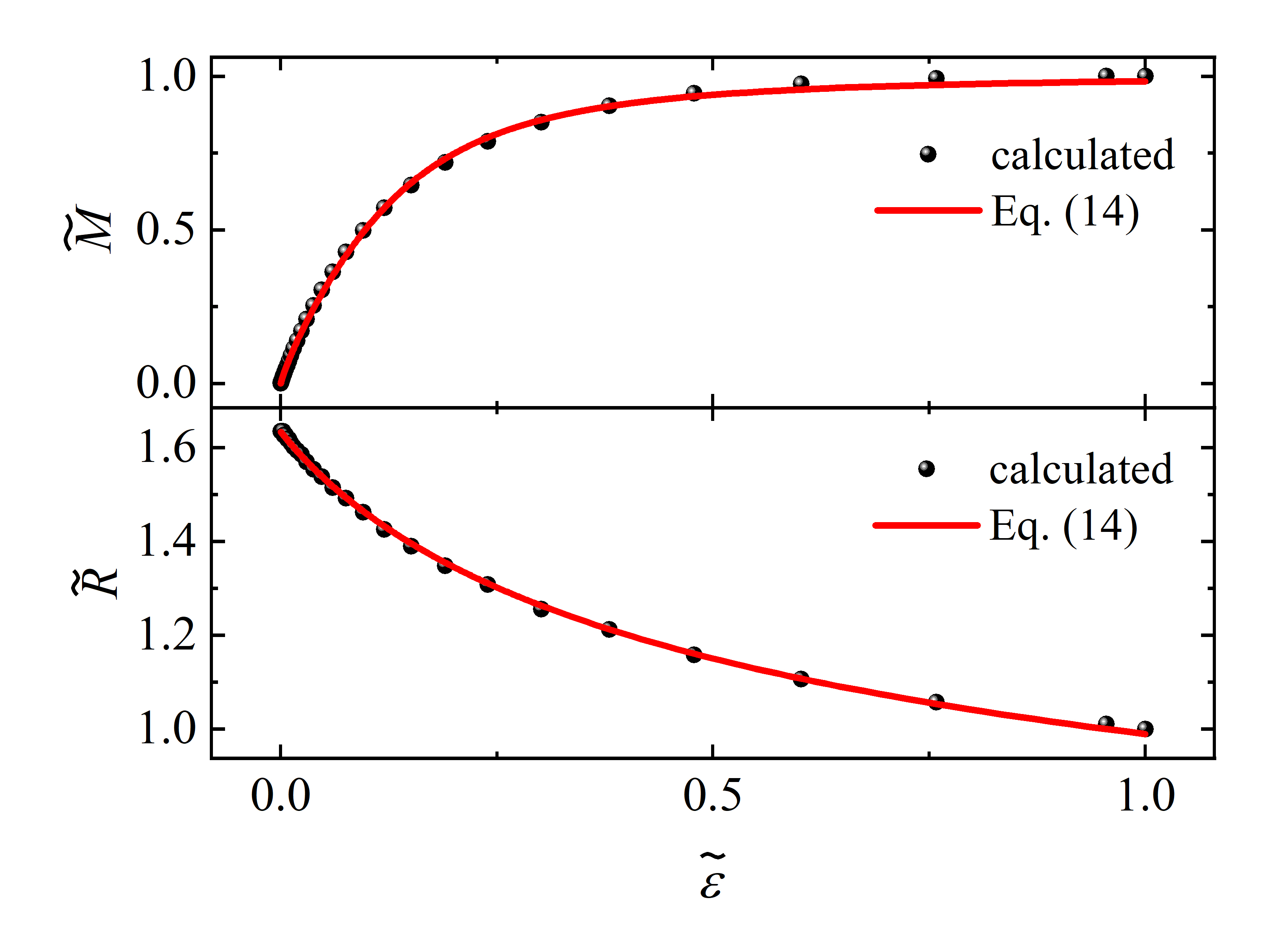}
    \caption{Dimensionless mass $\tilde{M}$ and radius $\tilde{R}$ as functions of dimensionless central density $\tilde{\varepsilon}$ on the stable branch. Scattered points are numerically calculated results from the TOV equations, and the red solid curves are given by Eq.~\eqref{eq:unified_fit}. For all parameter combinations ($m_\phi$ from $10^{-9}$ to $10^{3}$ GeV, $\lambda$ from $0.01\pi$ to $100\pi$), the calculated results are almost identical. Upper panel: $\tilde{M}$--$\tilde{\varepsilon}$; lower panel: $\tilde{R}$--$\tilde{\varepsilon}$.}
    \label{fig:MRE}
\end{figure}

\subsection{Universal Mass-Radius Relation}

After scaling transformation using the scaling relations \eqref{eq:Mmax_scaling} and \eqref{eq:Rmax_scaling} with the maximum mass $M_{\text{max}}$ and critical radius $R(M_{\text{max}})$, the dimensionless $M$--$R$ curves for all parameter combinations collapse onto a single curve. Quadratic polynomial fitting of the stable branch data yields:
\begin{equation}
\tilde{M} = b_0 + b_1 \tilde{R} + b_2 \tilde{R}^2,
\label{eq:MR_fit}
\end{equation}
with coefficients
\begin{equation}
b_0 = -1.9,\quad b_1 = 5.6,\quad b_2 = -2.7.
\label{eq:MR_coeffs}
\end{equation}

This fit has been verified over multiple independent parameter combinations covering $m_\phi$ from $10^{-9}$ to $10^{3}$ GeV and $\lambda$ from $0.01\pi$ to $100\pi$, spanning both weak and strong coupling regimes. The coefficients are very stable across the entire parameter space, with relative errors for all three parameters less than $3\%$. The $R^2$ for each set of parameters is $>0.9997$. This remarkable universality confirms the self-similar nature of boson star structure, suggesting an underlying mathematical structure worthy of further investigation.

Combining the scaling relations \eqref{eq:Mmax_scaling} and \eqref{eq:Rmax_scaling}, Eq.~\eqref{eq:MR_fit} enables one to directly obtain the radius of a boson star for any given mass on the stable branch without solving the TOV equations, simply by calculating
\begin{equation}
M_{\text{max}} = 0.1\,\frac{\sqrt{\lambda}}{m_\phi^2}\,M_\odot,\qquad
R(M_{\text{max}}) = 0.9\,\frac{\sqrt{\lambda}}{m_\phi^2}\ \text{km},
\end{equation}
and then solving for $\tilde{R}$ from $\tilde{M}$ using the inverse function of Eq.~\eqref{eq:MR_fit}.

Equation \eqref{eq:MR_fit} is valid within the range $\tilde{R}\lesssim1.5$ (i.e., $\tilde{M}\gtrsim0.5$), with a reproduction accuracy of $R^2 > 0.9995$ for the numerical results. The remarkable similarity of scaled stellar configurations across twelve orders of magnitude in boson mass and four orders of magnitude in self-coupling strength is an extraordinary discovery. It suggests that the universality found in this paper may extend beyond the specific quartic potential to a broader class of dark matter equations of state with similar scaling properties.

\begin{figure}[htbp]
    \centering
    \includegraphics[width=0.8\columnwidth]{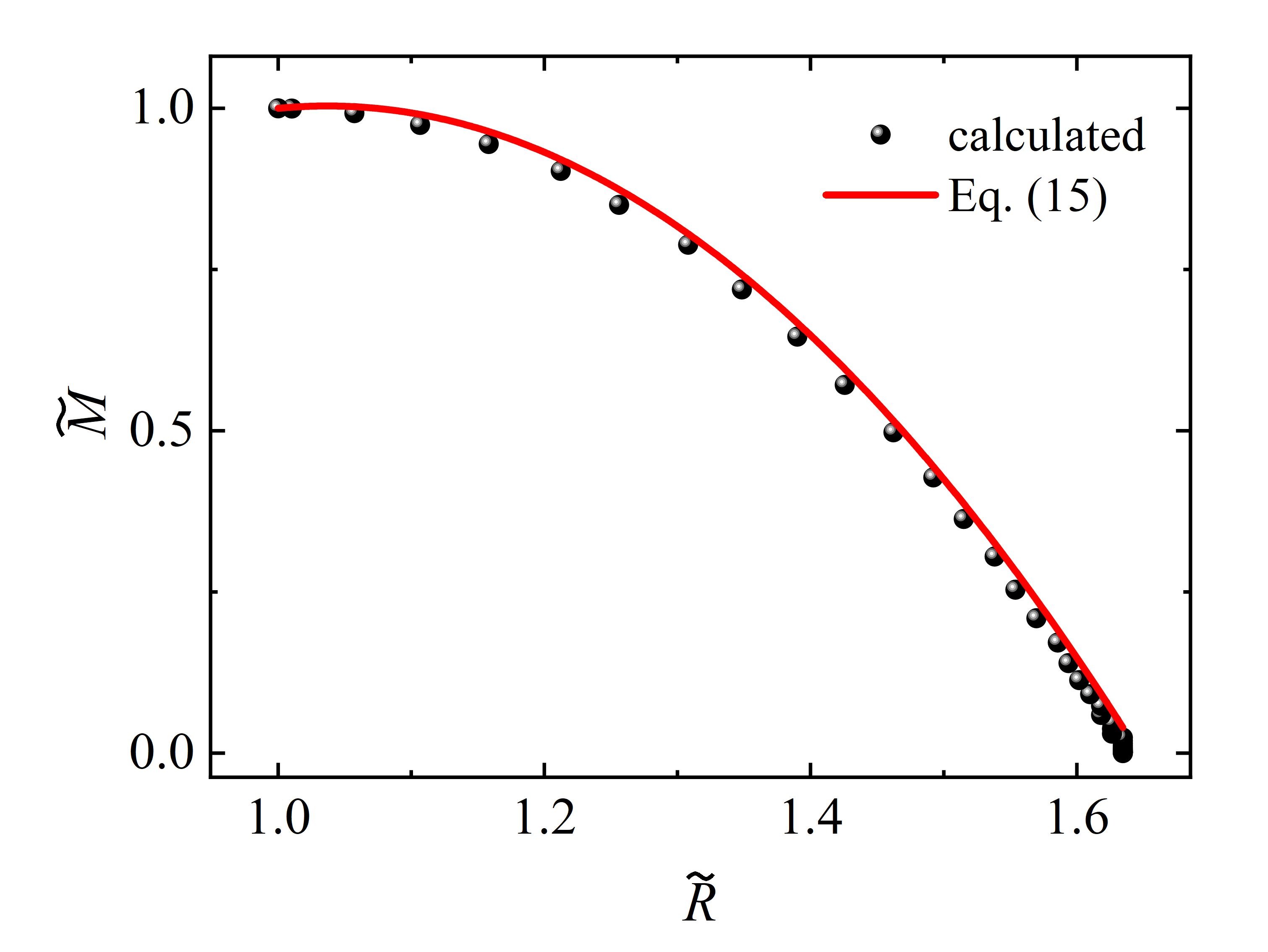}
    \caption{Dimensionless mass-radius relation on the stable branch. Scattered points are numerically calculated results from the TOV equations, and the red solid curve is given by Eq.~\eqref{eq:MR_fit}. For all parameter combinations ($m_\phi$ from $10^{-9}$ to $10^{3}$ GeV, $\lambda$ from $0.01\pi$ to $100\pi$), the calculated results are almost identical.}
    \label{fig:MR_fit}
\end{figure}

\subsection{Definition of the Central Density Range}

On the stable branch, as the central density decreases, the mass of the boson star decreases rapidly, and the gravitational binding weakens significantly. When $\tilde{\varepsilon} \ll 1$, the configuration mass is much smaller than $M_{\text{max}}$, and the radius approaches its maximum $R \approx 1.634 R(M_{\text{max}})$, with an extremely weak surface gravitational field. In traditional compact object (neutron star, white dwarf) research, attention is typically focused on configurations near the maximum mass on the stable branch, as such configurations have sufficiently strong gravitational binding to be observed through X-rays, gravitational waves, etc. \cite{Lattimer2001, Lattimer2004}. For example, reviews of neutron star mass-radius relations focus on configurations near the maximum mass \cite{Lattimer2007}. Based on the fitting accuracy analysis described above, this paper considers $\tilde{\varepsilon} \gtrsim 0.05$ (corresponding to $\tilde{M} \gtrsim 0.3$) as the central density range of primary astrophysical significance. Within this region, the accuracy of all analytical fitting formulas is better than $5\%$; for $\tilde{\varepsilon} < 0.05$, the boson star mass is less than $30\%$ of the maximum mass, the gravitational binding is significantly weakened, and it is difficult to distinguish from dark matter halos or ordinary gas clouds with current observational capabilities. Therefore, this paper does not take such configurations as the main research object.

\section{Boson Star Interpretation of Sgr A* at the Galactic Center}
\label{sec:sgra}

The compact radio source Sgr A* at the Galactic center is the closest massive compact object to Earth. Its mass has been precisely measured to be $(4.154 \pm 0.014) \times 10^6 M_\odot$ \cite{Gravity2022}, corresponding to a Schwarzschild radius of approximately $1.27 \times 10^7$ km. Observations by the Event Horizon Telescope further reveal the shadow structure of Sgr A*, with an angular diameter of $48.7 \pm 7$ $\mu$as, corresponding to a linear scale of approximately $6.6 \times 10^7$ km \cite{EHT2022a,EHT2022b}.

In this section, we use the scaling relations obtained in Section~\ref{sec:scaling} to test whether Sgr A* can be interpreted as a boson star with quartic self-interaction.

For a stable boson star, its total mass $M$ cannot exceed the maximum mass $M_{\text{max}}$, and its radius $R$ cannot be smaller than the critical radius $R(M_{\text{max}})$. Therefore, the observational requirements for Sgr A* are
\begin{align}
M_{\text{obs}} &\le M_{\text{max}}, \\
R_{\text{obs}} &\ge R(M_{\text{max}}),
\end{align}
where $M_{\text{obs}} = 4.154 \times 10^6 M_\odot$ and $R_{\text{obs}} = 6.6 \times 10^7$ km.

Substituting the scaling relations \eqref{eq:Mmax_scaling} and \eqref{eq:Rmax_scaling} into the above inequalities yields
\begin{align}
0.1 \frac{\sqrt{\lambda}}{m_\phi^2} &\ge 4.154 \times 10^6, \\
0.9 \frac{\sqrt{\lambda}}{m_\phi^2} &\le 6.6 \times 10^7.
\end{align}

Thus, $m_\phi$ must satisfy
\begin{equation}
\sqrt{\frac{0.9}{6.6 \times 10^7}} \lambda^{1/4} \le m_\phi \le \sqrt{\frac{0.1}{4.154 \times 10^6}} \lambda^{1/4},
\label{eq:constraint_general}
\end{equation}
We obtain
\begin{equation}
1.17 \times 10^{-4} \lambda^{1/4} \ \text{GeV} \le m_\phi \le 1.55 \times 10^{-4} \lambda^{1/4} \ \text{GeV}.
\label{eq:constraint_gev}
\end{equation}
Converting to keV units ($1 \ \text{GeV} = 10^6 \ \text{keV}$):
\begin{equation}
    117 \ \text{keV} \le \frac{m_\phi}{\lambda^{1/4}} \le 155 \ \text{keV}.
    \label{eq:constraint_kev}
\end{equation}

Equation~\eqref{eq:constraint_kev} follows from the extremal conditions $M_{\rm obs} \le M_{\max}$ and $R_{\rm obs} \ge R(M_{\max})$, which are necessary but not sufficient for the boson star interpretation---any viable configuration must simultaneously satisfy the $M$--$\varepsilon_0$ and $R$--$\varepsilon_0$ relations on the stable branch. A more stringent constraint is obtained by requiring that the observed mass and radius correspond to a specific point on the unified fitting formula \eqref{eq:unified_fit}. From Eqs.~\eqref{eq:unified_fit} and \eqref{eq:mass_radius_ratio}, the mass-to-radius ratio depends solely on the dimensionless central density $\tilde{\varepsilon}$:
\begin{equation}
\frac{M}{R} = \frac{M_{\max}}{R(M_{\max})}\,
\frac{\bigl[1+(5\tilde{\varepsilon})^{-2}\bigr]^{-0.42}}
{1.634\,\bigl[1+5\tilde{\varepsilon}\bigr]^{-0.28}}.
\label{eq:MR_ratio_fit}
\end{equation}
Substituting the observed values $M_{\rm obs}=4.154\times10^{6}\,M_\odot$ and $R_{\rm obs}=6.6\times10^{7}$\,km, together with $M_{\max}/R(M_{\max})=1/9\,M_\odot/{\rm km}$ from Eq.~\eqref{eq:mass_radius_ratio}, yields $\tilde{\varepsilon}\approx0.207$. The corresponding dimensionless mass is $\tilde{M}(0.207)\approx0.758$, from which $M_{\max}=M_{\rm obs}/\tilde{M}\approx5.48\times10^{6}\,M_\odot$. Combining this with the scaling relation $M_{\max}=0.1\sqrt{\lambda}/m_\phi^2$ from Eq.~\eqref{eq:Mmax_scaling} gives
\begin{equation}
\frac{m_\phi}{\lambda^{1/4}} \approx 1.35\times10^{-4}\ {\rm GeV}
= 135\ {\rm keV}.
\label{eq:constraint_refined}
\end{equation}
This value lies within the interval \eqref{eq:constraint_gev}, confirming the self-consistency of the two approaches.

\begin{figure*}
    \centering
    \includegraphics[width=0.8\textwidth]{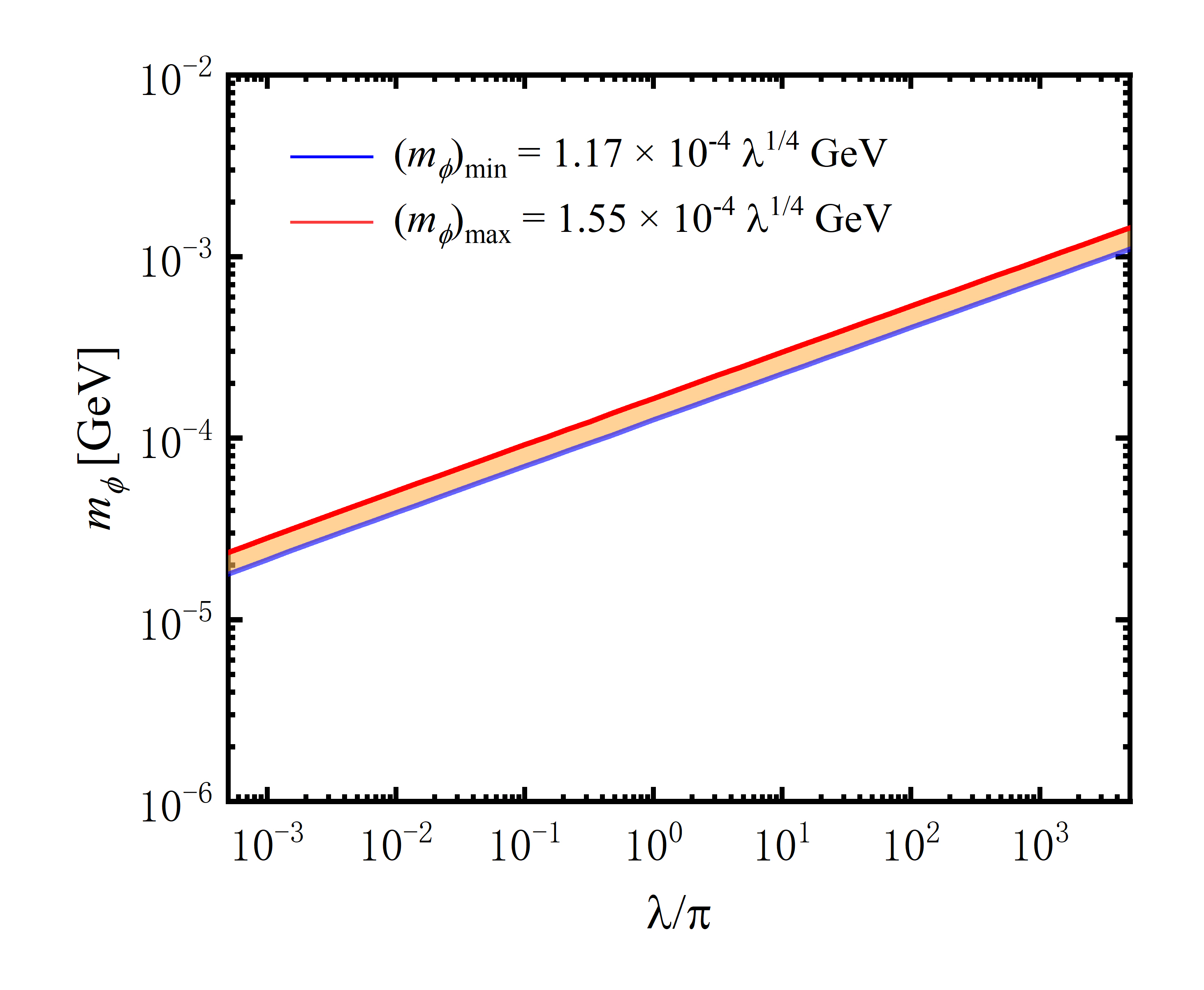}
    \caption{Allowed parameter space for Sgr A* as a bosonic dark matter star. The blue solid line is the lower bound from the radius constraint $(m_{\phi})_{min} = 1.17 \times 10^{-4} \ \lambda^{1/4}$ GeV, and the red solid line is the upper bound from the mass constraint $(m_{\phi})_{max} = 1.55 \times 10^{-4} \ \lambda^{1/4}$ GeV. The orange shaded region between the two boundaries is the parameter space satisfying the observed mass ($4.154\times10^6 M_\odot$) and shadow size ($6.6\times10^7$ km) of Sgr A*. Note that the combined parameter $m_\phi/\lambda^{1/4}$ is restricted to the narrow range $(1.17$--$1.55) \times 10^{-4}$ GeV.}
    \label{fig:parameter_space}
\end{figure*}

Figure~\ref{fig:parameter_space} displays the allowed parameter space from the bounding conditions. The blue solid line represents the lower bound from the radius constraint $(m_{\phi})_{\rm min} = 1.17 \times 10^{-4} \ \lambda^{1/4}$ GeV, and the red solid line represents the upper bound from the mass constraint $(m_{\phi})_{\rm max} = 1.55 \times 10^{-4} \ \lambda^{1/4}$ GeV. The orange shaded region between the two boundaries is the parameter space consistent with Sgr A* observations; the refined value $m_\phi/\lambda^{1/4}=1.35\times10^{-4}$\,GeV from Eq.~\eqref{eq:constraint_refined} lies near the center of this region.

\section{Summary}
\label{sec:SUMMARY}

This paper systematically investigates the structural properties of bosonic dark matter stars with a quartic self-interaction potential. Using the equation of state derived from the potential $V(\phi)=\frac{\lambda}{4}|\phi|^4$ and solving the TOV equations, we obtain precise scaling relations for the maximum mass, critical radius, and critical central density: $M_{\text{max}}=0.1\sqrt{\lambda}/m_\phi^2\,M_\odot$, $R(M_{\text{max}})=0.9\sqrt{\lambda}/m_\phi^2$ km, $\varepsilon_{\text{max}}=2.1\times10^5\,m_\phi^4/\lambda$ MeV/fm$^3$, where $m_\phi$ is in GeV. The ratio $M_{\text{max}}/R(M_{\text{max}})\approx 1/9\,M_\odot/\text{km}$ is a universal constant, verified across all parameter combinations. Based on these scaling relations, we construct global analytical fits for the entire stable branch: the $M$--$\varepsilon_0$ and $R$--$\varepsilon_0$ relations can be described by the unified dimensionless function $\tilde{Y}=A/[1+(5\tilde{\varepsilon})^h]^s$, with different $(A, h, s)$ parameter sets for $\tilde{M}$ and $\tilde{R}$; the dimensionless mass-radius curve collapses to a single universal quadratic relation $\tilde{M}=-1.9+5.6\tilde{R}-2.7\tilde{R}^2$. Applying these scaling relations to Sgr A* at the Galactic center,  we obtain the constraint $117\ \text{keV}\le m_\phi/\lambda^{1/4}\le 155\ \text{keV}$, indicating that the self-interacting bosonic dark  matter stars are viable candidates for Sgr A*.

This remarkable universality across twelve orders of magnitude in boson mass and four orders of magnitude in self-coupling strength suggests an underlying mathematical structure—likely rooted in the scaling symmetry of the Einstein-Klein-Gordon system—and may extend to a broader class of dark matter equations of state. The complete set of analytical formulas presented in this paper—the scaling relations for the maximum mass, critical radius, and critical central density, the unified fitting functions for $M(\varepsilon_0)$ and $R(\varepsilon_0)$, and the universal quadratic $M$--$R$ relation—enables researchers to quickly obtain boson star configurations for arbitrary parameter combinations without repeated numerical integration of the TOV equations. Future higher-resolution EHT observations, GRAVITY stellar orbit monitoring, and next-generation dark matter detection experiments will further test the boson star hypothesis and the scaling relations reported in this paper.

\begin{acknowledgments}
  This work is partly supported by the National Natural Science Foundation of China (No. 12475123, No. 12225504, No.12321005).
\end{acknowledgments}

\end{document}